\documentclass[preprint,aps,floats]{revtex4}
\usepackage{epsfig}
\usepackage{CJK}
\usepackage{graphicx}
\usepackage{dcolumn}
\usepackage{bm}
 \usepackage{multirow}
 \usepackage{array}
  \usepackage{ams}
\newcommand{\PreserveBackslash}[1]{\let\temp=\\#1\let\\=\temp}
\newcolumntype{C}[1]{>{\PreserveBackslash\centering}p{#1}}
\newcolumntype{R}[1]{>{\PreserveBackslash\raggedleft}p{#1}}
\newcolumntype{L}[1]{>{\PreserveBackslash\raggedright}p{#1}}
\def\lsim{\mathrel {\vcenter {\baselineskip 0pt \kern 0pt
    \hbox{$&lt;$} \kern 0pt \hbox{$\sim$} }}}
\def\gsim{\mathrel {\vcenter {\baselineskip 0pt \kern 0pt
    \hbox{$&gt;$} \kern 0pt \hbox{$\sim$} }}}
\newcommand{\U}{{\cal {U}}}
\begin{document}

\title{Large Dimuon Asymmetry In $B_s- \bar B_s$ Mixing \\From Unparticle Induced $\Gamma^{12}_s$}
\author{Bo Ren$^1$, Xiao-Gang He$^{1,2}$, Pei-Chu Xie$^1$}
\affiliation{
$^1$INPAC, Department of Physics, Shanghai Jiao Tong University, Shanghai, China\\
$^2$Department of Physics and Center for Theoretical
Sciences, National Taiwan University, Taipei, Taiwan\\
}

\date{\today}

\begin{abstract}
Exchange of unparticle stuff of dimension $d_\U$ with FCNC interaction
can induce $M^{12,u}$ and $\Gamma^{12,u}$ causing meson and
anti-meson mixing with the relation $\Gamma^{12,u}/M^{12,u} = 2
\tan(\pi d_\U)$. We show that this type of unparticle contribution
can provide the much needed large $\Gamma^{12}_s$ to explain the
recently observed anomalously large dimuon asymmetry in $B_s -\bar
B_s$ system reported by D0 collaboration. The same interaction can also accommodate
large mixing induced CP violation in $B_s \to J/\psi \phi$ indicated by CDF and D0 data. Experimental data can provide
constraints on the unparticle dimension and scale.
\end{abstract}


\maketitle


Recently the D0 Collaboration has reported evidence for an anomalously large CP violation in the like-sign
dimuon charge asymmetry \cite{d0} which is attributed to semileptonic decays of $b$ hadrons defined by
\begin{eqnarray}
A_{\rm sl}^b \equiv \frac{N_b^{++}-N_b^{--}}{N_b^{++}+N_b^{--}}\; ,
\end{eqnarray}
where $N_b^{++}\;(N^{--}_b)$ is the number of events with two $b$ hadrons ($b\; \bar b$) decaying semileptonically into
$\mu^+\mu^+ X\;(\mu^- \mu^-X)$. The D0 result \cite{d0}, $A_{\rm sl}^b = - (9.57 \pm 2.51 \pm 1.46)\times 10^{-3}$
with the first and the second errors being statistical and systematic ones,
is 3.2$\sigma$ away from the standard model (SM) prediction of $-0.2 \times 10^{-3}$ \cite{sm}.

$A_{\rm sl}^b$ is related to the ``wrong-charge'' asymmetries $a^{d,s}_{\rm sl}$ in $B_d$ and $B_s$ decays
\begin{eqnarray}
a_{\rm sl}^q \equiv \frac{\Gamma(\bar{B}_q \to \mu^+ X)-\Gamma(B_q \to \mu^-X)}{\Gamma(\bar{B}_q \to \mu^+ X)+\Gamma(B_q \to \mu^-X)} \;.
\end{eqnarray}
With known values for mixing parameters of $B_d$ and $B_s$ systems, one obtains~\cite{d0,PDG}
\begin{eqnarray}
A^b_{\rm sl}&=& (0.506\pm 0.043) a^d_{\rm sl} + (0.494\pm 0.043)a^s_{\rm sl}\;.
\end{eqnarray}

The above leads to~\cite{d0} $a^s_{\rm sl} = -0.0146\pm 0.0075$, after insert the known value~\cite{d0,hfag} $a_{\rm sl}^d = -0.0047\pm 0.0046$.
Combining direct information on $a_{\rm sl}^s$ from CDF  \cite{cdf}
and D0~\cite{d01}, one can extract an average value of
\begin{eqnarray}
(a_{\rm sl}^s)_{\rm ave} \approx - (12.7\pm 5.0) \times 10^{-3}\;.
\end{eqnarray}

The ``wrong-charge'' asymmetry for $B_s -\bar B_s$ mixing is determined by the matrix elements $M^{12}_s$ and $\Gamma^{12}_s$ in the Hamiltonian,
\begin{eqnarray}
a^s_{\rm sl} =  -{\textmd{Im}(M^{12}_s \Gamma^{12*}_s)\over
|M^{12}_s|^2 + |\Gamma^{12}_s|^2/4} \approx
- \frac{|\Gamma^{12}_s|}{|M_s^{12}|}\sin\phi_s\;. \label{as-f}
\end{eqnarray}
In the above, for the last step of approximation, we have neglected
small corrections of order $|\Gamma^{12}_s|^2/|M^{12}_s|^2$.


The SM prediction of $|\Gamma^{12,SM}_s|/|M_s^{12,SM}|$ is $(4.97\pm
0.94)\times 10^{-3}$ and the phase $\phi_s$ is small leading to a
value $(2.1\pm 0.6)\times 10^{-5}$ for $a^s_{\rm sl}$ \cite{sm}.
This is 2.5$\sigma$ away from $(a^s_{\rm sl})_{\rm ave}$, and the
predicted central value is more than two orders of magnitude
smaller. If the D0 result is confirmed, it is an indication of new
physics beyond the SM. Implications of this result have been studied
recently by several
groups~\cite{soni,Eberhardt,dighe,fox,chen,buras,Ligeti,pich,babu,dunn,dhg,Choudhury:2010ya,Chen:2010aq,Parry:2010ce,Ko:2010mn,Delaunay:2010dw,Bai:2010kf,Kubo:2010mh,
  Blum:2010mj,Wang:2010vv,Berger:2010wt,Dutta:2010ji,Lenz:2010gu,Oh:2010vc,Park:2010sg,Chao:2010mq,Buras:2010xj}.

Different values of $M^{12}_s$ and $\Gamma^{12}_s$ than those predicted by SM are needed. In
many models beyond the SM, there are new FCNC interactions which can
easily induce large modification to $M_s^{12}$, but not to
$\Gamma_s^{12}$, such as $Z'$ and SUSY models. If modifications to
$\Gamma^{12}_s$ is negligible, i.e. $\Gamma_s^{12} =
\Gamma^{12,SM}_s$, one can write~\cite{d0,sm}
\begin{eqnarray}
a^s_{\rm sl} =  - \frac{|\Gamma^{12,SM}_s|}{|M_s^{12, SM}|}\frac{\sin\phi_s}{|\Delta_s|} = - (4.97\pm 0.94)\times 10^{-3}\frac{\sin\phi_s}{|\Delta_s|}\;.
\label{as-f}
\end{eqnarray}
with
\begin{eqnarray}
M_s^{12} = M_s^{12,SM} + M^{12,NP}_s = M_s^{12,SM}\Delta_s = |M^{12,SM}_s| |\Delta_s|e^{i\phi_s}\;.\label{def-phase}
\end{eqnarray}
In the above, $M^{12,NP}_s$ indicates the contribution from beyond the SM new physics. We have neglected small SM contribution to $\phi_s$ and adopted the phase convention that $\Gamma^{12}_s$ is real.

Since the SM contribution to the phase $\phi_s$ is small, it is
convenient to write the expression explicitly in terms of the new
physics contribution defined by $M^{12,NP}_s/|M^{12,SM}_s| = R\,
\exp(-i\phi^{NP})$, one then has~\cite{dhg}
\begin{eqnarray}
&&a^s_{\rm sl} = {|\Gamma^{12,SM}_s|\over |M^{12,SM}_s|}{R\sin \phi^{NP}\over 1 + 2R\cos\phi^{NP} + R^2}\;,\nonumber\\
&&|\Delta_s| =\sqrt{1+2R\cos\phi^{NP} + R^2}\;.
\end{eqnarray}

The SM prediction for $\Delta M_s \approx 2 |M^{12}_s|$ agrees with data well, $|\Delta_s|$ is only allowed to vary in a limited region, $0.92\pm 0.32$ fixed by the experimental value  \cite{PDG}
$\Delta M_s = (17.77\pm 0.12)ps^{-1}$ and the SM prediction  \cite{sm} $(19.30\pm 6.74)ps^{-1}$. In order to reproduce the  D0
result, naively it seems that $\sin\phi_s$ would have to exceed the physical range  $|\sin \phi_s| < 1$ to reach the central value from data.
A large beyond SM new physics contribution $\Gamma^{12,NP}_s$ to $\Gamma^{12}_s$ is therefore called for. Several attempts have been made in this direction~\cite{dighe,dunn,dhg,Oh:2010vc}.

In this work we show that unparticle FCNC interaction can provide the much needed large $\Gamma^{12,u}_s$ to explain the observed anomalously large dimuon asymmetry in $B_s -\bar B_s$ system by D0 collaboration. Using experimental data, one can constrain the dimension $d_\U$ and the unparticle scale.


Several years ago Georgi proposed an interesting idea to describe
possible scale invariant effect at low energies by unparticles
\cite{Georgi:2007ek}. An unparticle operator $O_{\cal{U}}$ of
dimension $d_\U$ resulting from some scale invariant theory at high
energy may interact with SM particles at low energy in the following
form
\begin{eqnarray} \lambda \Lambda_{\cal{U}} ^{4-d_{SM} - d_\U}
O_{SM} O_{\cal{U}}.
\end{eqnarray}
where $O_{SM}$ is composed of SM fields.

If the $O_{SM}$ is a bi-quark operator with different quarks,
exchange of unparticle stuff can induce meson and antimeson
mixing~\cite{b-mixing}. Unlike usual tree level contributions to
meson oscillations from heavy particle exchange which produces a
small $\Gamma^{12}$, the unparticle may have sizeable contributions
to both $M^{12,u}$ and $\Gamma^{12,u}$ due to unparticle fractional
dimension $d_\U$ leading to a phase factor~\cite{Georgi:2007ek,
phase} $(-1)^{d_\U -2} = e^{-i\pi d_\U}$ in the propagator which is
proportional to $M^{12}-i\Gamma^{12}/2$. This leads to an
interesting relation $\Gamma^{12,u}/M^{12,u} = 2 \tan(\pi d_\U)$.
Depending on the dimension $d_\U$, $\Gamma^{12,u}$ can even be
larger than $M^{12,u}$.

Contributions to $M^{12,u}$ and $\Gamma^{12,u}$ for a class of
operators $O_{SM}$ with dimension less than or equal to 4
interacting with unparticles, respecting SM symmetries
\cite{operator}, have been studied in Ref.\,\cite{he}. Here we take
a scalar unparticle $O_\U$ interaction with quarks, respecting SM
gauge symmetry, of the following forms for illustration,
\begin{eqnarray}
i\tilde \lambda^D_{ij}\Lambda_{\cal{U}}^{-d_\U}\bar d^i_R  \gamma_\mu d^j_R \partial^\mu O_{\cal{U}}\;,\;\;\;\;i\tilde \lambda^Q_{ij}\Lambda_{\cal{U}}^{-d_\U}\bar Q^i_L  \gamma_\mu Q^j_L
\partial^\mu O_{\cal{U}}\;,
\end{eqnarray}
where $d^i_R$ and $Q^{i T}_L = (u^i_L, d^i_L)$ are the right-handed down-quarks and left-handed quarks, respectively.

After using equation of motion for quarks, one can write the above tree level FCNC interaction as
\begin{eqnarray}
&&{\cal L} = \Lambda_\U^{1-d_\U} {m_j\over \Lambda_\U} \bar d_i [\tilde \lambda^D_{ij} ({m_i\over m_j}  R  - L)  + \tilde \lambda^Q_{ij} ({m_i\over m_j}  L  - R)]d_j O_{\cal{U}}\;.
\end{eqnarray}
Let $j$ be the heavier and $i$ be the lighter quarks, the term proportional to $m_i/m_j$ can be neglected.

Note that the above Lagrangian, in general, can induce $B_s \to O_\U$ transition making a large contribution to $B_s$ invisible decay rate for $i=s$ and $j = b$. If there is no similar interaction for
$i =d$, this will also induce a significant lifetime difference between $B_d$ and $B_s$. These problems can be solved if the couplings $\tilde \lambda^D_{sb}$ and $\tilde \lambda^Q_{sb}$ are equal, $\tilde \lambda^D_{sb} = \tilde \lambda^Q_{sb}$, such that the above Lagrangian is proportional to $\bar s b O_\U$ which cannot induce $B_s \to O_\U$ transition. We will work with this condition.

Exchange of unparticle in the t and s channels as shown in Fig. 1,
we obtain the effective Hamiltonian $H^\U_{eff}$ responsible for
meson-antimeson mixing
\begin{eqnarray}
H^\U_{eff} = {A_{d_\U}\over 2 \sin(\pi
d_\U)}\Lambda_\U^{2(1-d_\U)}{m^2_j\over \Lambda_\U^2} e^{-i\pi
d_\U} {1\over
4} \left ( {1\over (s-\mu^2)^{2-d_\U}} + {1\over (t-\mu^2)^{2-d_\U}}\right
) \left (\bar d_i[\tilde \lambda^D_{ij}L +  \tilde \lambda^Q_{ij}R]  d_j\right )^2.\label{matrix}
\end{eqnarray}
Here $A_{d_\U} =
(16\pi^{5/2}/(2\pi)^{2d_\U})\Gamma(d_\U+1/2)/(\Gamma(d_\U
-1)\Gamma(2d_\U))$. We have used $(iA_{d_\U}/2\sin(\pi
d_\U))e^{-id_\U \pi} (1/(p^2-\mu^2)^{2-d_\U})$ for the scalar unparticle propagator~\cite{break}.
One can easily identify the
factor $e^{-i\pi d_\U}$ in the propagator producing a non-zero $\Gamma^{12,u}$ mentioned earlier.

In the above we have followed Ref.~\cite{break} to use a modified unparticle
propagator to take into account the breaking of conformal invariance at a scale $\mu$ lower than $\Lambda_\U$ by unparticle interaction with SM particles.
When $p^2$ becomes smaller than $\mu^2$, the unparticle contribution to $\Gamma_{12}$ vanishes.
The scale $\mu$ is not known, since we rely on the unparticle effect to generate $\Gamma^{12}_s$, we will allow it to be lower than $B_s$ mass.
One should not allow $\mu$ to be too small to allow
$b\to s O_\U$ process to happen which will contribute too large an invisible decay rate for $b\to s O_\U$ which changes $B_s$ lifetime dramatically.
For this reason we assume that $\mu$ is in the range of $m_{B_s} > \mu > m_{B_s} - m_K$ such that $b\to s O_\U$ is forbidden.

\begin{figure}[t!]
\includegraphics[width=5.5 in]{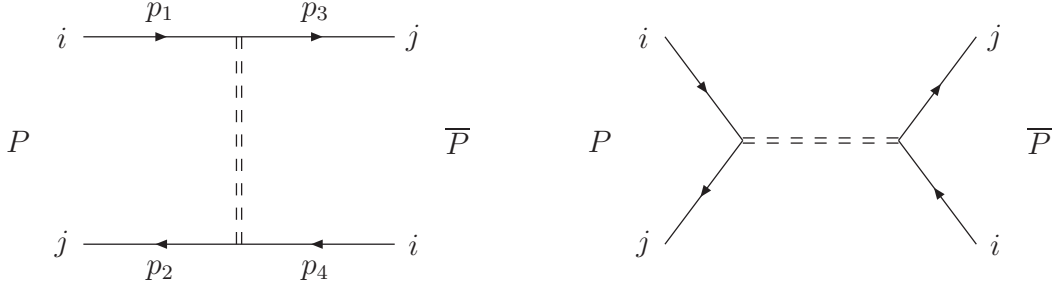}
\caption{\label{feynman} \small The $t$ and $s$ channel
contributions to meson-antimeson oscillation.}
\end{figure}

In $B_s-\bar B_s$ system, mesons are made of a $i=s$ light quark and a $j=b$ heavy.
In the heavy quark limit, one has $s = t\approx m^2_b \approx m^2_{B_s}$. With this
approximation and theoretical matrix elements for the relevant
operators, we have
\begin{eqnarray}
M^{12,u}_s- i {1\over 2} \Gamma^{12,u}_s &\approx & {A_{d_\U} \over \sin(d_\U \pi)}\left ({m^2_{B_s}-\mu^2\over \Lambda_\U^2 }\right
)^{d_\U-2} \left ({m^2_{B_s}\over \Lambda_\U^2 }\right
)^{2}{f^2_{B_s} \over 12 m_{B_s}}B_{B_s}
{e^{-i\pi d_\U}\over 8}\nonumber\\
&\times &\left ((\tilde \lambda^D_{sb})^2 + (\tilde \lambda^Q_{sb})^2 - 6((\tilde \lambda^D_{sb})^2 - (\tilde \lambda^Q_{sb})^2)
+ 2 \tilde \lambda^D_{sb}\tilde \lambda^Q_{sb}\right )\;,\label{ff}
\end{eqnarray}
where $B_{B_s}$ is the bag factor which is equal to 1 in the vacuum saturation and factorization approximation.
With the condition $\tilde \lambda^D_{sb} = \tilde \lambda^Q_{sb}$, the factor in the second line of the above equation becomes $4(\tilde \lambda^Q_{sb})^2$.

It is clear from Eq. (\ref{ff}) that because of the phase factor
$e^{-i\pi d_\U}$ in the propagator, there is a relation between
$M^{12,u}_s$  and $\Gamma^{12,u}_s$ that
\begin{eqnarray}
{\Gamma_s^{12,u}\over M_s^{12,u} } = 2\tan(\pi d_\U).
\end{eqnarray}

There may be a sizeable contribution to $\Gamma_s^{12,u}$ at tree
level which is not possible for usual tree level heavy  particle
exchange. For $d_\U$ equal to half integers, there is no
contribution to $M_s^{12,u}$, but there is for $\Gamma_s^{12,s}$.
It has been shown~\cite{he} that a large class of unparticle tree
level interaction contributions to meson and antimeson mixing can be
written in the form of Eq.~(\ref{um}).

One can group unknown parameters in such a way that
\begin{eqnarray}
M^{12,u}_s = |M^{12,u}_s|e^{-i\phi_\U}\;,\;\;\Gamma^{12,u}_s = 2|M^{12,u}_s|e^{-i\phi_\U} \tan(\pi d_\U)\;,
\label{um}
\end{eqnarray}
where
\begin{eqnarray}
|M^{12,u}_s|e^{-i\phi_\U} = |A_{d_\U}  \left ({m^2_{B_s}\over \tilde
\Lambda_\U^2}\right )^{d_\U} {f^2_{B_s} \over 24 m_{B_s}}
 B_{B_s} \cot(d_\U \pi)|e^{-i\phi_\U}\;.\label{mr}
\end{eqnarray}
Here we have identified the phase of $\tilde \lambda^Q_{sb}$ to be $- \phi_\U/2$ and defined the reduced unparticle scale $\tilde \Lambda_\U = \Lambda_\U (1-\mu^2/m^2_{B_s})^{1/d_\U-1/2}|\tilde \lambda^Q_{sb}|^{-1/d_\U}$. $\tilde \Lambda_\U$ is an effective indicator of the scale of unparticle physics. This quantity actually
contains three unknown quantities, $\Lambda_\U$, $|\lambda^Q_{sb}|$ and $\mu$. Since we do not have good ideas about their individual sizes, we will use the reduced scale for discussions later. In general $\tilde \lambda^Q_{sb}$ is complex, therefore $\phi_\U$ can take an arbitrary value between 0 to $2\pi$.

Combining the SM contribution and neglect small phase there, we have
\begin{eqnarray}
&&a^s_{\rm sl} = {|\Gamma^{12,SM}_s|\over |M^{12,SM}_s|}{R\sin(\phi_\U)\over 1+2 R\cos(\phi_\U)+R^2}
- {2 \tan(d_\U \pi) R\sin(\phi_\U)\over 1+2 R\cos(\phi_\U) + R^2}\;,\nonumber\\
&&|\Delta_s| = \sqrt{1+2R\cos(\phi_\U) + R^2}\;.\label{master}
\end{eqnarray}
The first term in the expression for $a^s_{\rm sl}$ is similar to the
contribution without modification to $\Gamma^{12}_s$ analyzed in
Ref. \cite{dhg}. The second term is new which comes from unparticle
modification to $\Gamma^{12}_s$.

There are two constraints with three parameters, $R$, $\phi_\U$ and $y = \tan(d_\U \pi)$.
It is not possible to determine all the theoretical parameters. However, one can express two of the theoretical parameters as a function
of the experimental measurable quantities and another theoretical parameter. We find
\begin{eqnarray}
&&\cos(\phi_\U) = {|\Delta_s|^2-1-R^2\over 2R}\;,\nonumber\\
&&y = {1\over 2} \left ({|\Gamma^{12,SM}_s|\over |M^{12,SM}_s|} \pm {2 a^s_{\rm sl} R |\Delta_s|^2\over \sqrt{4R^2-(|\Delta_s|^2 -1-R^2)^2}}\right )\;.
\end{eqnarray}
The ``$\pm$'' sign in the expression for $y$ comes from the need of taking a square root for $\sin(\phi_\U)$ from knowing $\cos(\phi_\U)$.

Since the physical range of $\cos(\phi_\U)$ is limited to be from -1 to 1, from the first equation about, one can obtain a
constrain for the allowed ranges for
$R$. We show this in Fig. 2 for several values of $|\Delta_s|$. In the plot we have used the central values for the relevant SM quantities.

With the allowed range for $R$ known, for given $a^s_{\rm sl}$ and $|\Delta_s|$, $y$ can be determined as a function of $R$.
For example, with central values for $a^s_{\rm sl}$ and $|\Delta_s|$, for $R = 0.5, 1.0, 1.5$, we have
$y=0.008, 0.009, 0.012$  and $y =-0.003, -0.004, -0.007$ with ``-'' and``$+$'' signs in eq. 18, respectively.
$a^s_{\rm sl}$ and $|\Delta_s|$ can be accommodated with reasonable values for
other parameters. $y$ can take both positive or negative signs. To have further information on $d_\U$, one should also consider $d_\U = n + (\mathrm{arc}{\tan{y}})/\pi$ with n an arbitrary integer.  However, if one considers
the relation between the reduced scale $\tilde \Lambda_\U$ and
$d_\U$, one would prefer not too large a $d_\U$. If one limits
$d_\U$ between 1 and 2, when y takes a positive value, one should take
$n=1$. When $y$ is negative, it is reasonable to take $n =2$.

\begin{figure}[t!]
\includegraphics[width=3.5 in]{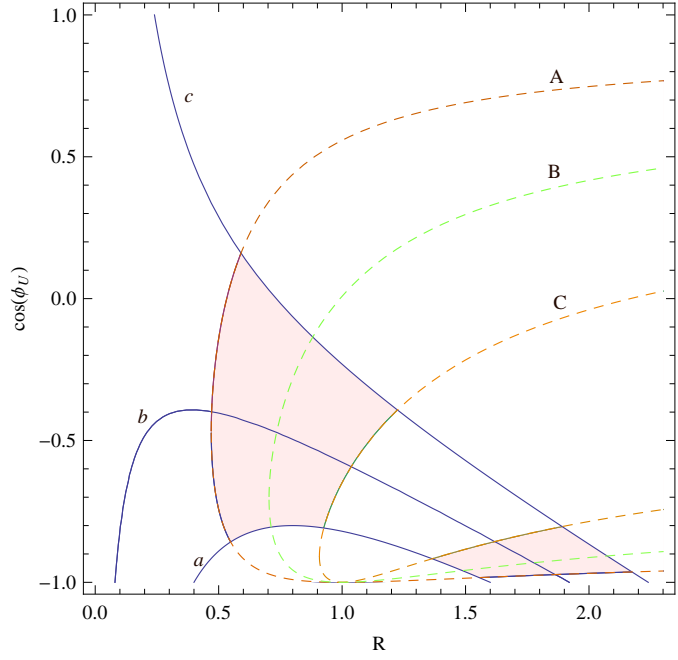}
\caption{\label{feynman} \small $ R$ vs. $\cos(\phi_\U)$. The solid and dashed curves are for
$|\Delta_s|$ and $\sin \phi_s$ taking their central and one $\sigma$ upper and lower values, respectively.
The shaded areas are the one $\sigma$ overlap regions. ($|\Delta_s|=0.6$ (a), $|\Delta_s|=0.92$ (b), $|\Delta_s|=1.24$ (c);
$\sin \phi_s=-0.470$ (A), $\sin \phi_s=-0.704$ (B)), $\sin \phi_s=-0.908$ (C)).}
\end{figure}

\begin{figure}[t!]
\includegraphics[width=4.0 in]{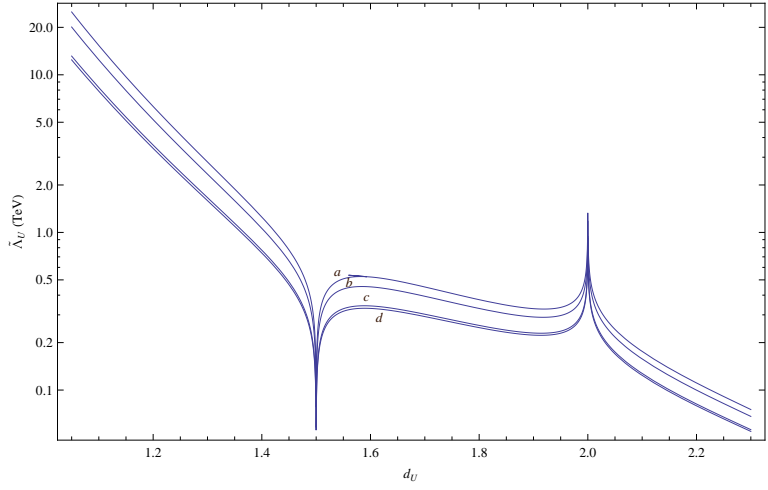}
\caption{\label{feynman} \small $\tilde \Lambda_\U$ vs. $d_\U$ for
fixed $R$ = 0.46; 0.73; 1.78; 2.00, in order from up to down (magenta
(a); green (b); blue (c); red (d)), respectively.}
\end{figure}

The unparticle interaction we have introduced can also have effects on other flavor changing processes which may exclude some of the parameter space
favored by producing the $a^s_{\rm sl}$ value obtained by D0. For $\Delta S = 1$ processes, such as $B_{s,d}$ decays with SM final
states and unparticle as intermediate state, there is the need to introduce new unparticle interaction
with other SM particles, one can adjust the new couplings to be sufficiently small to satisfy bounds from experimental data.
This type of interaction will not affect the analysis carried out earlier and we will not discuss them further. There are, however,
experimental observables which depend on the quantities we have discussed. For example mixing induced CP violation in $B_s$ decays can be affected by modifications to $M^{12}_s$, in particular, experimental data on mixing induced CP violation in $B_s \to J/\psi \phi$ can further constraint the interaction discussed.

Recently analysis from CDF and D0 data show that there are indications of mixing induced CP violation in $B_s\rightarrow J/\psi\phi$ decay with~\cite{Aaltonen}
\begin{eqnarray}
&&\beta^{J/\psi\phi}_s = (0.39^{+0.18}_{-0.14})\cup(1.18^{+0.14}_{-0.18}),
\end{eqnarray}
where the second set in the last line is just the complement of $\pi/2$. This reflects the ambiguity in the determination of $\beta^{J/\psi\phi}_s$.

If there is no new interaction in the decay amplitude, $-2\beta^{J/\psi \phi}_s$ is the phase of $M^{12}_s$.
The SM prediction for $\beta^{J/\psi\phi}_s$ is given by~\cite{sm}, $\beta^{J/\psi\phi(SM)}_s = \mathrm{Arg}(-\frac{V_{cb}V_{cs}^*}{V_{tb}V_{ts}^*})\approx0.019\pm0.001$.
The central value of $\beta^{J/\psi \phi(SM)}_s$ is much smaller than that from CDF and D0 and can be neglected. If confirmed, this also requires new physics beyond SM.
Even though we do not introduce new unparticle interactions in the decay amplitude, $M^{12}_s$ is modified which can affect mixing induced CP violation in $B_s \to J/\psi \phi$. We have $-2 \beta^{J/\psi\phi}_s \approx \phi_s$. Here $\phi_s$ is defined in Eq. (\ref{def-phase}). The two values for $\beta^{J/\psi\phi}_s$ give approximately the same $\sin(\phi_s)$ with the central values given by $-0.704$ and the one $\sigma$ range of $-0.908 \sim -0.470$.

Translating into $R$ and $\phi_\U$ defined earlier, we have
\begin{eqnarray}
\sin \phi_s = \frac{-R\sin(\phi_\U)}{\sqrt{1+2 R\cos(\phi_\U)+R^2}}.
\end{eqnarray}
It is clear that a non-zero $\phi_\U$ can modify the SM prediction for $\beta^{J/\psi \phi}_s$.

Using the above one can further
constraint the allowed regions for $R$ and $\cos (\phi_\U)$. From the relation between $\phi_s$ and $\phi_\U$, one finds $\sin(\phi_\U)$ must be positive.
The allowed regions for $\cos(\phi_\U)$ are shown in Fig. 2. The shaded
regions are the one $\sigma$ allowed common regions.


With three unknowns and three input experimental data points, we can completely express
$R$, $\cos(\phi_\U)$ and $y$ as functions of  $|\Delta_s|$, $\phi_s$ and $a^s_{\rm sl}$. We have
\begin{eqnarray}
&&R =\sqrt{1\pm2|\Delta_s|\cos\phi_s+|\Delta_s|^2}\;,\nonumber\\
&&\cos(\phi_{\U}) = \frac{-1\mp|\Delta_s|\cos \phi_s}{\sqrt{1\pm2|\Delta_s|\cos \phi_s+|\Delta_s|^2}}\;,\\
&&y = {1\over 2} \left ( {|\Gamma^{12,SM}_s|\over |M^{12,SM}_s|}+{a^s_{\rm sl}|\Delta_s| \over \sin \phi_s}\right )\;.\nonumber
\end{eqnarray}

We now use the above equations to determine central values and their one $\sigma$ errors for $R$, $\cos(\phi_\U)$ and $y$. Since the error correlations between
the three experimental observable quantities are not know, a complete analysis is not possible. To have some idea about the ranges for the theoretical parameters, we will assume that they are independent quantities satisfying Gaussian distribution and take the average one $\sigma$ lower and upper bounds as their one $\sigma$ errors for illustration. The results are listed in
Table 1. For a given set of input data, there are two solutions for $R$ and $\cos(\phi_\U)$, but the solution for $y$ is single valued. We see that $y$ is restricted to be positive now.
The large dimuon asymmetry in $B_s - \bar B_s$ mixing and large mixing induced CP violation in $B_s \to J/\psi \phi$ can be simultaneously explained.

\begin{table}{
\begin{tabular}{|c|c|c|}
  \hline
  $R$ & $\cos(\phi_{\U})$ & $y$ \\
  \hline
  {$1.776\pm0.185$}&{$-0.931\pm0.055$}&\multirow{2}{*}{$0.011\pm0.004$}\\\cline{1-2}
  {$0.735\pm0.276$}&{$-0.472\pm0.208$}&{}\\
  \hline
\end{tabular}}\caption{The central values and one $\sigma$ bounds of $R$, $\cos(\phi_{\U})$ and $y$. }
\end{table}

Finally, we briefly comment on the implications of constraints on the effective unparticle scale $\tilde \Lambda_\U$.
Using the results obtained on $R$, we can work backward to obtain
information about the reduced unparticle scale, $\tilde \Lambda_\U$.
For a given $R$, $\tilde \Lambda_\U$ can be expressed as a function
of $d_\U$. We show the results in Fig.~3. In plotting Fig.~3, we
have used $B_{B_s} =1$ and $f_{B_s} = 0.260~\textmd{GeV}$~\cite{Okamoto:2005zg}. We
can see, as expected for larger $d_\U$, a lower scale is needed
which may have already reached by current colliders. Since no
evidence of unpaticle effects have shown up explicitly, one should
set a lower limit for that. For a lower bound of $\tilde \Lambda_{\U}$ larger than 1
TeV, we see that $d_{\U}$ is restricted to be lower than 1.5.

\newpage


In summary we have shown that unparticle induced contribution to $B_s - \bar B_s$ mixing can provide the much needed large $\Gamma^{12}_s$ to explain the recently observed anomalously large dimuon asymmetry by D0 collaboration. The same interaction can also accommodate large mixing induced CP violation in $B_s \to J/\psi \phi$ from CDF and D0 data. Experimental data also provide constraints on unparticle dimension and scale parameters.

\vskip 1.0cm \noindent {\bf Acknowledgments}$\,$ We thank A. Lenz for some discussions. The work of
authors was supported in part by the NSC, NCTS and NNSF.

\vskip 1.0 cm \noindent
{\bf Note Added}

While we are in the final stage of finishing our work, a paper~\cite{jplee} by J. P. Lee appeared where unparticle effects on $B_s - \bar B_s$
was also discussed in the context of large dimuon asymmetry reported by D0. We do not agree Lee's interpretation for the part proportional to $\sin(\pi d_\U)$ generated due to the $e^{-i\pi d_\U}$ factor in the propagator. It contributes to
$\Gamma^{12}_s$, but not $M^{12}_s$. Our emphasis is on how the unparticle effects modification for $\Gamma^{12}_s$ plays a crucial role in explaining the data reported by D0.
In general a CP violating phase, in our notation $\phi_\U$,
can also exist. This phase is not considered in the analysis by Lee.


\begin{references}

\bibitem{d0} V.~M.~Abazov {\it et al.}  [D0 Collaboration],
  Phys.\ Rev.\  D {\bf 82}, 032001 (2010)
  [arXiv:1005.2757 [hep-ex]];
 V.~M.~Abazov {\it et al.}  [D0 Collaboration],
  Phys.\ Rev.\ Lett.\  {\bf 105}, 081801 (2010)
  [arXiv:1007.0395 [hep-ex]].


\bibitem{sm} A. Lenz and U. Nierste, J. High Energy Phys. {\bf 0706},
072(2007).

\bibitem{PDG} Particle Data Group, C. Amsler, et al., Phys. Lett. B {\bf 667}, 1 (2008), and 2010 edition.


\bibitem{hfag} E. Barberio et al. (HFAG), [arXiv:0808.1297 [hep-ex]].

\bibitem{cdf} CDF Collaboration, Note 9015, Oct. 2007.

\bibitem{d01} V.M. Abzov et al. [D0 Collaboration], [arXiv:0904.3907 [hep-ex]].








\bibitem{soni} A.~Soni, A.~K.~Alok, A.~Giri, R.~Mohanta and S.~Nandi,
 [arXiv:1002.0595 [hep-ph]].

\bibitem{Eberhardt} O.~Eberhardt, A.~Lenz and J.~Rohrwild,
  [arXiv:1005.3505 [hep-ph]].


\bibitem{dighe} A.~Dighe, A.~Kundu and S.~Nandi,
  Phys.\ Rev.\  D {\bf 82}, 031502 (2010)
  [arXiv:1005.4051 [hep-ph]].


\bibitem{fox} B.~A.~Dobrescu, P.~J.~Fox and A.~Martin,
  Phys.\ Rev.\ Lett.\  {\bf 105}, 041801 (2010)
  [arXiv:1005.4238 [hep-ph]].


\bibitem{chen} Chuan Hung Chen, Gaber Faisel,
e-Print: [arXiv:1005.4582 [hep-ph]].
\bibitem{buras} A. Buras, S. Gori, M. Carlucci and G. Isidori, e-print: [arXiv:1005.5310 [hep-ph]].


\bibitem{Ligeti}
  Z.~Ligeti, M.~Papucci, G.~Perez and J.~Zupan,
  [arXiv:1006.0432 [hep-ph]].
\bibitem{pich} M. Jung, A. Pich and P. Tuzon, e-print: [arXiv:1006.0470 [hep-ph]].

\bibitem{babu}
K.~S.~Babu and J.~Julio,
  [arXiv:1006.1092 [hep-ph]].

\bibitem{dunn} C. Bauer and N. Dunn, e-print: [arXiv:1006.1629 [hep-ph]].

\bibitem{dhg} N. G. Deshpande, Xiao Gang He and German Valencia, Phys. Rev. D {\bf82}, 056013 (2010)  [arXiv:1006.1682 [hep-ph]].


\bibitem{Choudhury:2010ya}
  D.~Choudhury and D.~K.~Ghosh,
  [arXiv:1006.2171 [hep-ph]].



\bibitem{Chen:2010aq}
  C.~H.~Chen, C.~Q.~Geng and W.~Wang,
  [arXiv:1006.5216 [hep-ph]].



\bibitem{Parry:2010ce}
  J.~K.~Parry,
  [arXiv:1006.5331 [hep-ph]].

\bibitem{Ko:2010mn}
  P.~Ko and J.~h.~Park,
  [arXiv:1006.5821 [hep-ph]].

\bibitem{Delaunay:2010dw}
  C.~Delaunay, O.~Gedalia, S.~J.~Lee and G.~Perez,
  [arXiv:1007.0243 [hep-ph]].

\bibitem{Bai:2010kf}
  Y.~Bai and A.~E.~Nelson,
  [arXiv:1007.0596 [hep-ph]].

\bibitem{Kubo:2010mh}
  J.~Kubo and A.~Lenz,
  [arXiv:1007.0680 [hep-ph]].



\bibitem{Blum:2010mj}
  K.~Blum, Y.~Hochberg and Y.~Nir,
  [arXiv:1007.1872 [hep-ph]].


\bibitem{Wang:2010vv}
  R.~M.~S.~Wang, Y.~G.~S.~Xu, M.~L.~S.~Liu and B.~Z.~S.~Li,
  [arXiv:1007.2944 [hep-ph]].

\bibitem{Berger:2010wt}
  C.~Berger and L.~M.~Sehgal,
  [arXiv:1007.2996 [hep-ph]].

\bibitem{Dutta:2010ji}
  B.~Dutta, Y.~Mimura and Y.~Santoso,
  [arXiv:1007.3696 [hep-ph]].

\bibitem{Lenz:2010gu}
  A.~Lenz {\it et al.},
  [arXiv:1008.1593 [hep-ph]].

\bibitem{Oh:2010vc}
  S.~Oh and J.~Tandean,
  [arXiv:1008.2153 [hep-ph]].



\bibitem{Park:2010sg}
  S.~C.~Park, J.~Shu, K.~Wang and T.~T.~Yanagida,
  [arXiv:1008.4445 [hep-ph]].

\bibitem{Chao:2010mq}
  W.~Chao and Y.~C.~Zhang,
 [arXiv:1008.5277 [hep-ph]].

\bibitem{Buras:2010xj}
  A.~J.~Buras,
  [arXiv:1009.1303 [hep-ph]].









\bibitem{Georgi:2007ek}
  H.~Georgi,
  Phys.\ Rev.\ Lett.\  {\bf 98}, 221601 (2007)
  [arXiv:0703260 [hep-ph]];
  H.~Georgi,
  Phys.\ Lett.\  B {\bf 650}, 275 (2007)
  [arXiv:0704.2457 [hep-ph]];
   K.~Cheung, W.~Y.~Keung and T.~C.~Yuan,
  Phys.\ Rev.\ Lett.\  {\bf 99}, 051803 (2007)
  [arXiv:0704.2588 [hep-ph]].




\bibitem{b-mixing}
  M.~Luo and G.~Zhu,
  [arXiv:0704.3532 [hep-ph]];
   C.~H.~Chen and C.~Q.~Geng,
  [arXiv:0705.0689 [hep-ph]];
  X.~Q.~Li and Z.~T.~Wei,
  Phys.\ Lett.\  B {\bf 651}, 380 (2007)
  [arXiv:0705.1821 [hep-ph]];
  R.~Mohanta and A.~K.~Giri,
  [arXiv:0707.1234 [hep-ph]];
  A.~Lenz,
  Phys.\ Rev.\  D {\bf 76}, 065006 (2007)
  [arXiv:0707.1535 [hep-ph]];
  G.~J.~Ding and M.~L.~Yan,
  [arXiv:0705.0794 [hep-ph]];
  Y.~Liao,
  Phys.\ Rev.\  D {\bf 76}, 056006 (2007)
  [arXiv:0705.0837 [hep-ph]];
  C.~D. Lu, W. Wang and Y.~M. Wang, Phys. Rev. D {\bf 76}, 077701(2007);
  C.~S.~Huang and X.~H.~Wu,
  [arXiv:0707.1268 [hep-ph]];
  R.~Mohanta and A.~K.~Giri,
  Phys.\ Rev.\  D {\bf 76}, 057701 (2007)
  [arXiv:0707.3308 [hep-ph]].

\bibitem{phase}
  C.~H.~Chen and C.~Q.~Geng,
  Phys.\ Rev.\  D {\bf 76}, 036007 (2007)
  [arXiv:0706.0850 [hep-ph]];
   C.~H.~Chen and C.~Q.~Geng,
  [arXiv:0709.0235 [hep-ph]].

\bibitem{operator}
  S.~L.~Chen and X.~G.~He, Phys. Rev. D {\bf 76}, 091702 (2007)
 [arXiv:0705.3946 [hep-ph]].


\bibitem{he} S.~L.~Chen et al., Eur. Phys. J. C {\bf59}, 899 (2009).

\bibitem{break}
P.~J.~Fox, A.~Rajaraman and Y.~Shirman,
  Phys.\ Rev.\ D {\bf 76}, 075004 (2007)
  [arXiv:0705.3092 [hep-ph]].




\bibitem{Aaltonen} T. Aaltonen et al. [CDF Collaboration], Phys. Rev. Lett. {\bf 100}, 161802 (2008);
                   V. M. Abazov et al. [D0 Collaboration], Phys. Rev. Lett. {\bf 101}, 241801 (2008).

\bibitem{Okamoto:2005zg}
  M.~Okamoto,
  PoS {\bf LAT2005}, 013 (2006)
  [arXiv:hep-lat/0510113].

\bibitem{jplee} Jong Phil Lee, Phys.\ Rev.\ D {\bf 82}, 096009 (2010) [arXiv: 1009.1730 [hep-ph]].


\end{references}
\end{document}